\documentclass[10pt]{iopart}
\usepackage{amsthm}
\usepackage{amssymb}
\usepackage{setstack}
\usepackage{iopams}
\usepackage{graphicx}

\def\Th{\Theta}

\def\sig{\sigma}

\def\3nab{\tilde{\nabla}}

\def\hsp5{\hspace{5mm}}
\newcommand{\sfrac}[2]{{\textstyle{#1\over#2}}}
\def\case#1/#2{\textstyle\frac{#1}{#2}}

\def\be {\begin{equation}}
\def\ee {\end{equation}}
\def\ber {\begin{eqnarray}}
\def\eer {\end{eqnarray}}
\def\bea {\begin{eqnarray}}
\def\eea {\end{eqnarray}}

\def\bc {\begin{center}}
\def\ec {\end{center}}

\def\etal\;{{\it et al.}}

\begin{document}

\title{A dynamical system approach to higher order gravity (Talk given at IRGAC 2006, July 2006) }

\author{\underline{Sante Carloni}$^{1}$,  Peter K. S. Dunsby$^{1,2}$}

\address{$1.$ Department of Mathematics and Applied Mathematics, \\
University of Cape Town, 7701 Rondebosch, South Africa,}
\address{$2.$ South African Astronomical Observatory, Observatory 7925,
Cape Town, South Africa.}

\begin{abstract}
The dynamical system approach has recently acquired great importance
in the investigation on higher order theories of gravity. In this
talk I review the main results and I give brief comments on the
perspectives for further developments.
\end{abstract}

\maketitle

\section{Introduction}
Since the very first proposal of General Relativity alternative
formulations of the law of gravity have been proposed for various
purposes: from unification of the fundamental interaction to the
explanation of the dark energy phenomenon. In spite of the great
variety of proposals, all these attempts present a common drawback:
they are much more complicated than General Relativity. This means
that it is very difficult to devise tests on these theories, and
hence there are not many opportunities to gain insights on the their
physics.

Obtaining such information is critical to the debate about the
nature of gravitation and its relation with the dark energy issue.
Particularly important in these respect are the indirect methods,
i.e. methods that allows to solve indirectly the gravitational field
equations or to obtain a qualitative idea of how these solutions
behave.

One very interesting methods of this type is based on the
application of the dynamical system approach to cosmology. This
method has been developed for general relativity since the 70s, and
it has lead to beautiful insights into the evolution of the
anisotropic cosmologies \cite{ellisbook}.

In this paper we summarize the results obtained from the application
of dynamical system theory to fourth order gravity and in particular
to $R^n$-gravity. We also discuss very briefly some further
applications of this method.

Unless otherwise specified, we will use natural units
($\hbar=c=k_{B}=8\pi G=1$) throughout the paper and Greek indices
run from 0 to 3. The semicolon represents the usual covariant
derivative and the ``dot" corresponds to time differentiation.
\section{$R^{n}-gravity$}
The action for the gravitational interaction in this theory reads
\begin{equation}\label{71-curv1}
{\cal A}=\int d^4x \sqrt{-g} \left[\chi(n) R^{n}+{L}_{M} \right]\;,
 \end{equation}
where $\chi(n)$ is a positive function of $n$ that reduces to $1$
for $n=1$. For $R\neq0$, the field equations for this theory can be
written as
\begin{eqnarray}\label{equazioni di campo Rn Rdiv0}
 \nonumber&&G_{\mu\nu}=  T_{\mu\nu}^{M}+ T_{\mu\nu}^{R}
=\frac{\tilde{T}_{\mu\nu}^{M}}{n\chi(n)
R^{n-1}}+\frac{1}{2n}g_{\mu\nu}(1-n)R
+\left[(n-1)\frac{R^{;\alpha\beta}}{R}
\right.\\
&&\left.+(n-1)(n-2)\frac{R^{;\alpha}
R^{;\beta}}{R^{2}}\right](g_{\alpha\mu}g_{\beta\nu}
-g_{\alpha\beta}g_{\mu\nu})\;,
\end{eqnarray}
where $\tilde{T}_{\mu\nu}^{M}$ is the stress energy tensor for the
standard matter. In this way, the non-Einsteinian part of the
gravitational interaction can be modelled as an effective fluid
which, in general, presents thermodynamic properties different from
standard matter \footnote{Note that the nature of the matter term
and the indetermination in the sign of $R$ makes this theory to be
fully meaningful only if we consider in the following only values of
$n$ belonging to the set of the relative numbers $\mathcal{Z}$ and
the subset of the rational numbers ${\mathcal{Q}}$, which can be
expressed as fractions with an odd denominator.}. In the following
we will analyze this equation in the
Friedmann-Lema\^{i}tre-Robertson-Walker (FLRW) and  Bianchi I
metrics with the aid of the dynamical system approach.

\section{Dynamical analysis of the FLRW case}
In the FLRW metric, the (\ref{equazioni di campo Rn Rdiv0}) take the
form
\begin{eqnarray}\label{Raychaudhuri Rn}
\nonumber &&2n\frac{\ddot{a}}{a}+n(n-1)H\frac{\dot{R}}{R}
+n(n-1)\frac{\ddot{R}}{R}+n(n-1)(n-2)\frac{\dot{R}^{2}}{R^{2}}-(1-n)\frac{R}{3}\\
&&+\frac{\mu}{3n\chi(n)R^{n-1}}(1+3w)=0\;,
\end{eqnarray}
\begin{equation}\label{friedmann Rn}
H^{2}+\frac{\kappa}{a^{2}}+H\frac{\dot{R}}{R}(n-1)-\frac{R}{6n}(1-n)-\frac{\mu}{3n\chi(n)R^{n-1}}=0\;.
\end{equation}
\begin{equation}\label{conservazione Rn}
\dot{\mu}+3H\mu (1+w)=0 \;.
\end{equation}
with
\begin{equation}\label{R FLRW}
R=-6\left(\frac{\ddot{a}}{a}+\frac{\dot{a}^{2}}{a^{2}}+
\frac{\kappa}{a^{2}}\right)
\end{equation}
where $H=\dot{a}/a$, $\kappa$ is the spatial curvature index and we
have assumed standard matter to be a perfect fluid with a barotropic
index $w$. Note that in the above equations we have considered $a$
and $R$ as two independent fields so that the equation are of
effective order two and the conservation equation for matter is the
same as standard GR \cite{RnSante}.

The form of the above equations suggests the following choice of
expansion normalized variables:
\begin{eqnarray}\label{dyn variables lin matt}
\fl x=\frac{\dot{R}}{R H}(n-1)\;,\ \ \ y=\frac{R}{6nH^{2}}(1-n)\;,\
\ \
 z=\frac{\mu}{3n \chi(n) H^{2} R^{n-1}}\;,\ \ \
K=\frac{\kappa}{a^{2}H^{2}}\;,
\end{eqnarray}
Differentiating (\ref{dyn variables lin matt}) with respect to the
logarithmic time $\mathcal{N}=\ln a$, we obtain the autonomous
system (matter is considered a perfect fluid)
\begin{eqnarray}\label{sistema compatto materia}
\nonumber x'=\frac{2(n-2)y}{n-1}-2 x-2x^2-\frac{x y}{n-1}+(1+x)z-3zw\;,\\
y'=\frac{y}{n-1}\left[(3-2n)x-2y+2(n-1)z+2(n-1)\right]\;,\\
\nonumber z'=z\left(2z-1-3x-\frac{2y}{n-1}-3w\right)\;,
1+x-y-z+K=0\;.
\end{eqnarray}
Note that the two planes $y=0$ and $z=0$ correspond to two invariant
submanifolds. This implies that for this system no finite global
attractor exists. The behavior of the scale factor corresponding the
fixed points can be found using the equation
\begin{equation}\label{H-puntifissi matt}
    \dot{H}=\left(x_{C}+\frac{y_{C}}{n-1}-z_{C}-1\right)H^{2}\;.
\end{equation}
If $n\neq1$ and $(n-1)(x_{C}-z_{C}-1)-y_{C}\neq0$ this equation can
be integrated to give
\begin{equation}\label{fattore di scala critico materia}
    a=a_{0}(t-t_{0})^{\alpha} \qquad\mbox{with}\qquad \alpha=\left(1-x_{C}-\frac{y_{C}}{n-1}+z_{C}\right)^{-1}\;.
\end{equation}
In the same variables the energy density $\mu$ can be written as
\begin{equation}\label{rho-dyn var}
    \mu=z y^{n-1}H^{2n}\;,
\end{equation}
thus for $n>1$ both the $(x,y)$ and $(x,z)$ planes are invariant
vacuum manifolds, but if $n<1$ the vacuum submanifold is not
necessarily compact.

A detailed analysis of this dynamical system is shown elsewhere
\cite{RnSante}. Here we will  focus on the interval $1.367\lesssim
n<2$. This interval is suggested by a fitting of the data coming
from WMAP and Supernovae Ia data \cite{Rnobs}.

\subsubsection{The vacuum Case}
In a vacuum spacetime (i.e. $\mu=0$), the variable $z$ is
identically zero and the third equation of (\ref{sistema compatto
materia}) becomes an identity. Let us analyze this case first.
Setting $x'=0$ and $y'=0$, we obtain the four fixed points shown in
Table \ref{tavola punti fissi vuoto 1}.
\begin{table}[t]
\caption{Coordinates of the  finite and asymptotic fixed points and
solutions for FLRW $R^{n}$-gravity in vacuum.} \centering
\bigskip
\begin{tabular}{cccc}
\br Point& Coordinates  & Solution/Behaviour \\ \mr
$\mathcal{A}$ & $[x,y]=[0,0]$ & $a=a_{0}t$\\
$\mathcal{B}$ & $[x,y]=[-1,0]$& $a= a_{0}(t-t_{0})^{1/2}$ (only for $n=3/2$) \\
$\mathcal{C}$ &$[x,y]=\left[\frac{2 (n-2)}{2 n-1}\;,\frac{4 n-5}{2
n-1}\right]
$& $a=a_{0}\;t^{\frac{(1-n)(2n-1)}{n-2}}$\\
$\mathcal{D}$ &$[x,y]=\left[2(1-n)\;, 2(n-1)^{2}\right]$& $
a=a_{0}t$
\\\br
$\left(\mathcal{A}_{\infty},\;\mathcal{C}_{\infty},\;\mathcal{D}_{\infty},\;\mathcal{F}_{\infty}\right)$
& $\theta=\left(0,\;\pi/2,\;\pi,\;3\pi/2\right)$& $|{\cal{N}}-{\cal N}_{\infty}|=\left[C_{1}\pm\frac{C_{0}}{2}(t-t_{0})\right]^{2}$\\
$\left(\mathcal{B}_{\infty},\;\mathcal{E}_{\infty}\right)$
&$\theta=\left(\pi/4,\;5\pi/4\right)$ & $ |{\cal N}-{\cal
N}_{\infty}|= \left[C_{1}\pm C_{0}\left|\frac{n-1}{2n-1}\right|
(t-t_{0})\right]^{\frac{2n-1}{n-1}}$\\\br
\end{tabular}
\label{tavola punti fissi vuoto 1}
\end{table}
In our interval the solution associate with the point $\mathcal{C}$
is a power law inflation and the only finite attractor. This is an
interesting result because this model contains cosmic histories that
naturally approach to a phase of accelerate expansion. However, the
presence of the invariant submanifolds already mentioned makes this
attractor not global and we have to check if there are other
attractors in asymptotic regime.

The idea that the system above might have a nontrivial asymptotic
structure comes from the fact that the phase plane is not compact.
The asymptotic analysis can be easily performed using the
Poincar\'{e} approach \cite{lefschetz}. We obtained six fixed points
which are summarized with their behavior in Table \ref{tavola punti
fissi vuoto 1}. The stability analysis shows that in the interval of
$n$ we have chosen there is another attractor (this time global):
the point $\mathcal{D}_{\infty}$. Using the asymptotic limit of
Equation \ref{H-puntifissi matt} (see \cite{RnSante}), we found that
$\mathcal{D}_{\infty}$ is associated with a Lema\^{i}tre type
evolution in which the universe reaches a maximum size and then
recollapses. The presence of $\mathcal{D}_{\infty}$ reduces the
measure of the set of initial conditions for which an orbit will
approach to $\mathcal{C}$, or in simpler words it makes this type of
cosmic histories ``less probable". A pictorial representation of the
whole (compactified) phase space is given in Figure \ref{figBI}.

\subsubsection{The non-vacuum Case}

When matter is present we consider the full system (\ref{sistema
compatto materia}). Setting $x'=0$, $y'=0$, $z'=0$ and solving for
$x,y,z$ we obtain seven fixed points and two fixed subspaces. The
first four points lay on the plane $z=0$ and correspond to the
vacuum fixed points. Of the other three fixed points, $\mathcal{G}$
is definitely the most interesting because of its associated
solution: $\displaystyle{a=a_{0} t^{\frac{2n}{3(1+w)}},\;\;
\mu=\mu_{0}t^{-2n}}$. The presence of such a fixed point brings the
idea that there could be cosmic histories in this model in which a(n
unstable) Friedmann-like phase is followed naturally by a phase of
accelerated expansion. Applying the standard tools of the dynamical
system, we discover that this is actually the case in our interval
for $n$. In fact, for $1.367\lesssim n<2$ the point $\mathcal{C}$ is
an attractor and the point $\mathcal{G}$ is a saddle, they are both
physical and placed in a connected sector of the phase space. Thus
in principle a cosmic history like the one pictured above is
possible. This is also confirmed by numerical investigation of the
dynamical system. The last thing to check is the presence of other
attractors in the asymptotic regimes. The compactification is
achieved in the same way as the vacuum case, but the equations
obtained are much more complicated. Here we will limit ourselves to
say that there are other attractors in the phase space, in the form
of subspaces and single points and their presence reduce the measure
of the set of initial conditions that lead to a cosmic history
connecting $\mathcal{G}$ and $\mathcal{C}$.

\section{Dynamical analysis of the LRS Bianchi I cosmologies }

In order to obtain the simplest possible form for the field
equations in the LRS Bianchi I metric, we use the 1+3 covariant
approach to cosmology \cite{ellisrev}. In this formalism the
cosmological equations can be written as
\begin{equation}
\dot{\Th} + \sfrac{1}{3}\,\Th^{2} + 2\sigma^2 -\frac{1}{2n}R -
(n-1)\frac{\dot{R}}{R}\Th+\frac{\mu}{\chi n R^{n-1}} =0,
\label{Ray:R^n_B1}
\end{equation}
\begin{equation}\label{3R:Rn_B1}
\sfrac{1}{3}\,\Th^{2}-\sigma^2+(n-1)\frac{\dot{R}}{R}\Th-\frac{(n-1)}{2n}R-\frac{\mu}{\chi
n R^{n-1}}=0\;,
\end{equation}
\begin{equation}\label{sigdot:Rn_LRS_B1}
\dot{\sig}= -\left(\Th+ (n-1)\frac{\dot{R}}{R}\right)\sigma \;,
\end{equation}
together with the (\ref{conservazione Rn}). Here $\Theta$ is the
volume expansion $\Theta =3\sfrac{\dot{a}}{a}=3H$ and $\sig$ is the
square root of the magnitude the symmetric shear tensor $\sig_{ab}$.
The set of expansion normalized variables is:
\begin{eqnarray}\label{DS:var}
\fl   x = \frac{3\dot{R}}{R\Th}(n-1)\; , \ \ \ \ y =
\frac{3R}{2n\Th^2}(n-1)\; , \ \ \ \  z = \frac{3\mu}{\chi
nR^{n-1}\Th^2}\;,\ \ \ \Sigma =\frac{3\sigma^2}{\Th^2}\; ,
\end{eqnarray}
and the dynamical system can be written as
\begin{eqnarray}\label{DS:eqn_mat2}
\Sigma'=-2\left[\left(\frac{2n-1}{n-1}\right)\;y+z\right]\Sigma, \nonumber \\
 y'=\frac{y}{n-1}\left[(2n-1)\Sigma-(2n-1)y+z+(4n-5)\right], \\
 z' =
z\left[(2-3w)-z+\Sigma-\left(\frac{3n-1}{n-1}\right)\;y\right],
\nonumber\\
1-\Sigma + x - y - z =0\nonumber .
\end{eqnarray}
The solution associated to every fixed point can be found via the
equations
\begin{eqnarray}
\fl\dot{\Th}=\left(\frac{n}{n-1}y_i-\Sigma_i-2\right)\frac{\Th^2}{3}\;,\qquad\qquad
\dot{\sigma}= -\frac{1}{3}(2+\Sigma_i+y_i+z_i)\;\Th\;\sigma.
\end{eqnarray}
For $n \neq 1$, $\sigma\neq 0$ and a R.H.S. different from zero,
these equations admit the solution
\begin{eqnarray}\label{DS:sol_gen_mat}
a=a_0\left(t-t_0\right)^\alpha, &&\qquad\alpha=\left(2+\Sigma_i-\sfrac{n}{n-1}y_i\right)^{-1}\;,\\
\sigma=\sigma_0 a^{\beta}, && \qquad \beta=-(2+\Sigma_i+y_i+z_i)\;.
\end{eqnarray}
As in the previous section we will focus on a specific intervals of
$n$ referring the reader to \cite{jannie} for a general analysis.
\subsubsection{The vacuum Case}
\begin{figure}
\begin{minipage}{5.5cm}
\includegraphics[width=5cm]{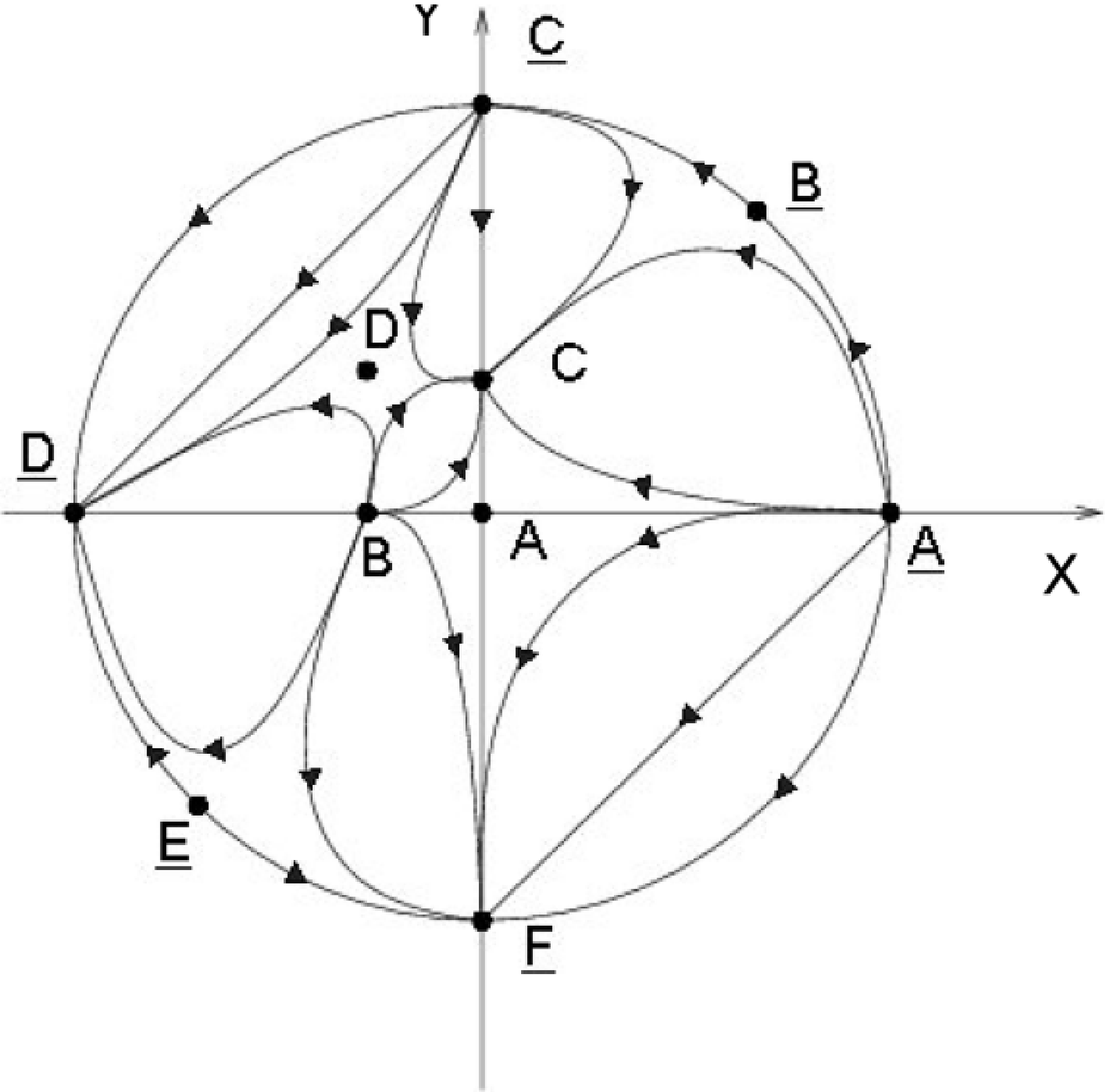}
\end{minipage}
\begin{minipage}{4cm}
\includegraphics[width=3cm]{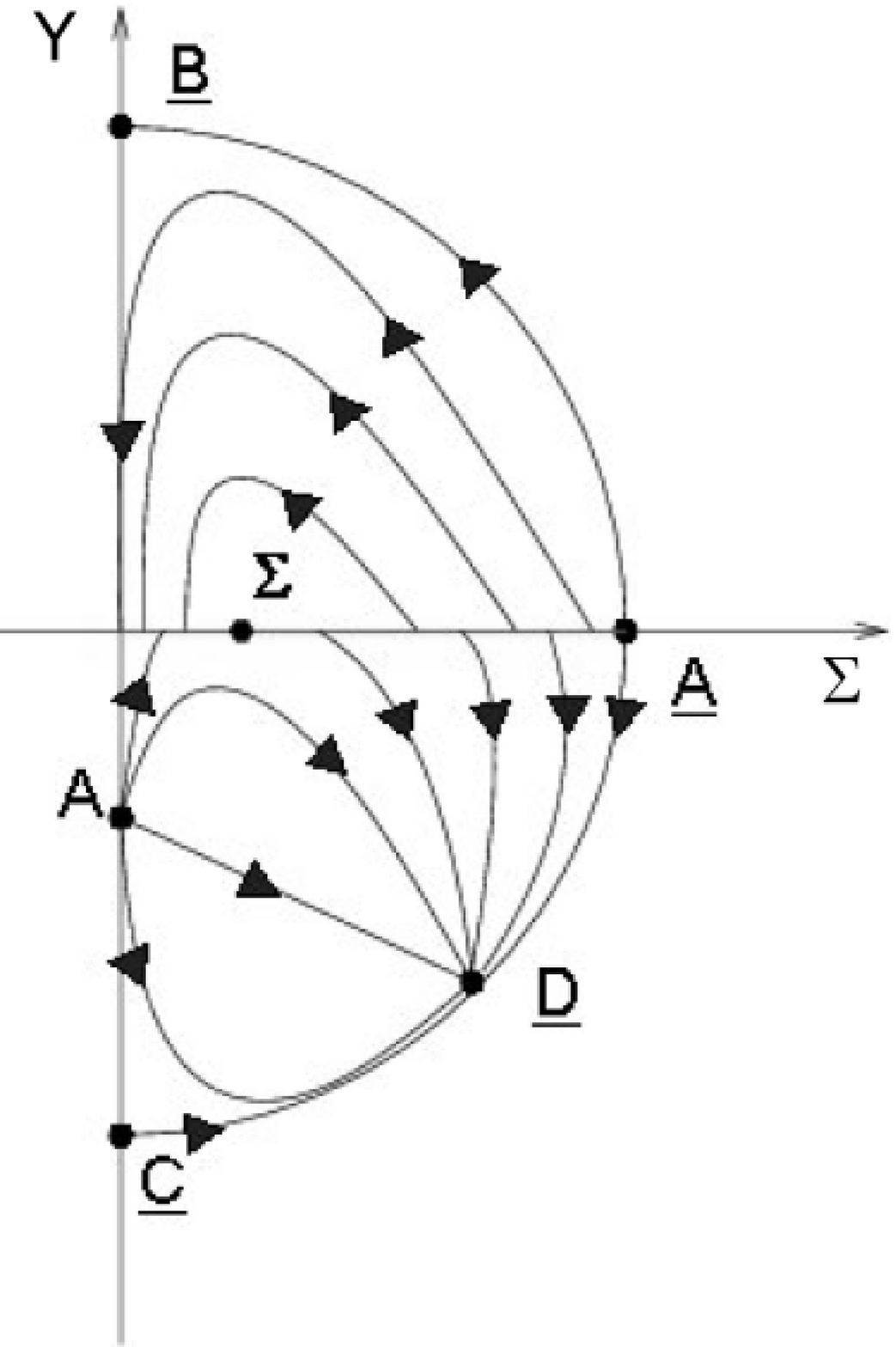}
\end{minipage}
\begin{minipage}{2cm}
\includegraphics[width=3cm]{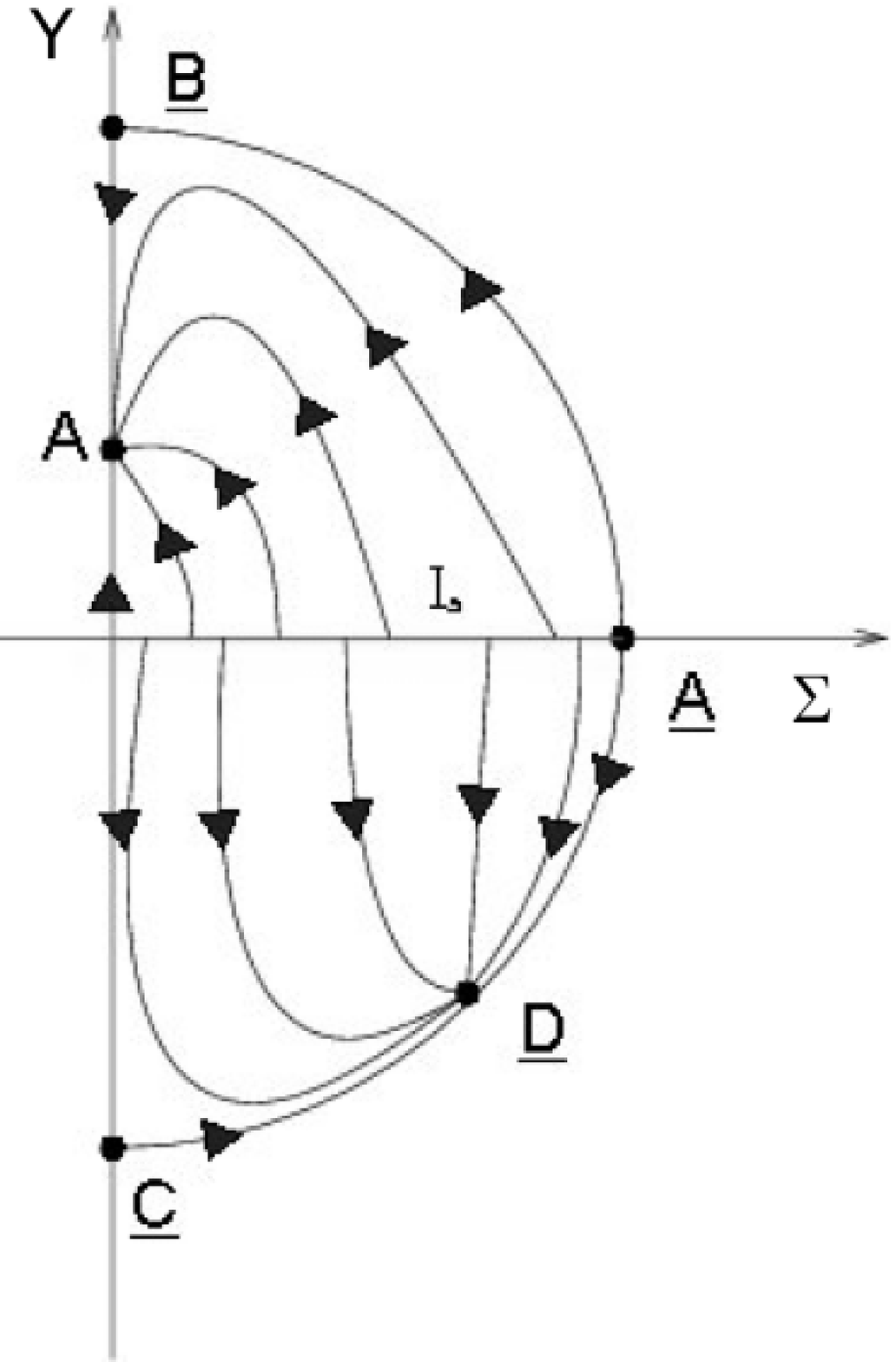}
\end{minipage}
\caption{Global (compactified) phase spaces for $R^{n}$-gravity. The
first picture represent the FLRW case with $1.36\lesssim n<2$; the
second the Bianchi I case for $1<n<5/4$ and the third the Bianchi I
$n>5/4$ case. The standard capital letters
 represent the finite fixed points the underlined capital
 letters represent the asymptotic fixed points. The point $\Sigma$
is nor a fixed point but it is defined as the last attractive point
of the fixed line. The curly capital ``l" represents the line
$\mathcal{L}_1$.}\label{figBI}
\end{figure}

In order to treat the vacuum case, we set $z=0$ and we neglect the
equation for $z$ as in the FLRW case. Setting $\Sigma'=0$ and $y'=0$
we obtain the fixed subspaces and their solutions (see Table
\ref{Table:vac_eigen}).
\begin{table}[tbp] \centering
\caption{The  finite an asymptotic fixed points and eigenvalues for
$R^n$-gravity in a LRS Bianchi I model.}
\begin{tabular}{ccccc}\\
\br  Fixed subspaces& Coordinates & Scale Factor& Shear \\ \mr
$\tilde{\mathcal{A}}$ &
$(\Sigma,y)$=$\left(0,\sfrac{4n-5}{2n-1}\right)$ &
$a=a_0\left(t-t_0\right)^{\sfrac{(1-n)(2n-1)}{(n-2)}}$ & $\sigma=0$
\\ \mr
$\mathcal{L}_1$ & $(\Sigma,y)$=$\left(\Sigma_*,0\right)$&
$a=a_0\left(t-t_0\right)^{\sfrac{1}{2+\Sigma_*}}$,  &
$\sigma=\sigma_0 a^{-(2+\Sigma_*)}$ \\\br $\mathcal{A}_\infty$ &
$\phi$=$0$ &$\tau-\tau_{\infty}=\frac{1}{\Sigma_c}\ln
\left|t-t_0\right|+C$ &
$\sigma=\sigma_0 a^{-(2+\Sigma_c)}$ \\
$\left(\mathcal{B}_\infty,\; \mathcal{C}_\infty\right)$ &
$\phi$=$\left(\frac{\pi}{2}\;,\frac{3\pi}{2}\right) $ &
$|\tau-\tau_\infty|=\left[C_1 \pm C_0\left|\sfrac{n-1}{2n-1} \right|(t-t_0)\right]^\frac{2n-1}{n-1}$ & $\sigma=0$ \\
$\mathcal{D}_\infty$ & $\phi$=$\sfrac{7\pi}{4}$ &
$|\tau-\tau_\infty|=\left[C_1 \pm C_2(t-t_0)\right]^2$ & $\sigma=\sigma_0$ \\
\br
\end{tabular}\label{Table:vac_eigen}
\end{table}
Note that the fixed point $\tilde{\mathcal{A}}$ represents the same
solution associated to the point $\mathcal{A}$. Direct verification
with the cosmological equations reveals that this point is
anisotropic i.e. $\sig_{ab}=0$. The presence of such a point is
particularly interesting because it means that there could be cosmic
histories in which there is an isotropic state for an otherwise
anisotropic universe. Two interesting cases arise. When $1<n<5/4$,
the point $\tilde{\mathcal{A}}$ is a repeller and part of the orbits
emerging from it approach the fixed line. This means that in these
cosmic histories the universe starts in an isotropic state and then
develops anisotropies (see figure \ref{figBI}). The presence of an
isotropic past attractor is not present in GR and implies that in
$R^n$-gravity, like in brane cosmology, there is no need for special
initial conditions for inflation to begin \cite{jannie}. When
$n>5/4$, $\tilde{\mathcal{A}}$ is an (isotropic) attractor and all
the orbits above the fixed line start on $\mathcal{L}_1$ and
converge to it (see Figure \ref{figBI}). This scenario presents the
same smooth transition between a decelerated and accelerated
expansion of the FLRW case.
If we analyze  the asymptotic regime for this last case we can see
that for  $n>5/4$ the point $\mathcal{D}_\infty$ is an attractor
(see Table \ref{Table:vac_eigen}), but is in a separate section of
the phase space with respect to $\tilde{\mathcal{A}}$, sot that
$\tilde{\mathcal{A}}$ can be considered an ``effective" global
attractor.

\subsubsection{The Matter Case}
Setting $\Sigma'=0$, $y'=0$ and $z'=0$ in (\ref{DS:eqn_mat2}) we
obtain, together with the vacuum fixed points, other two fixed
points, one of which ($\tilde{\mathcal{G}}$) is an isotropic point
associated with the same solution of $\mathcal{G}$. Let us check if
it is possible to have a cosmic history similar to the one we found
in FLRW. For $n>5/4$, the fixed line contains repulsive points
nearby the origin, the point $\tilde{\mathcal{G}}$ is unstable and
the point $\tilde{\mathcal{A}}$ in an attractor. Since these points
are not separated by any invariant subspace, there is in principle
an orbit that connects them. Along this orbit, the universe would
start in an anisotropic state, evolve towards a more isotropic state
to smoothly approach to a phase of accelerated expansion. Of course
the issue of determining how many other attractors are present (and
so how ``probable" this evolution is) is still present and we have
to perform a detailed asymptotic analysis in order to understand
this point. Using the results of \cite{jannie} it is possible to
show that there are other attractors in the phase space that might
influence the global evolution. However, the outcome is that an
orbit like the one described above is still possible and surely
deserves more study.

\section{Conclusion}
In this paper we hope to have shown how useful and powerful the
dynamical system approach is in dealing with complicated model like
fourth order gravity ones. The application of this method has
allowed new insights on higher order cosmological models and has
shown a deep connection between these theories and the cosmic
acceleration phenomenon, which is worth to be further studied.

The perspective for future application of the dynamical system
approach to higher order gravity can be basically divided in two
thrusts. The first one is to analyze more complicated Lagrangians.
Some of work in this direction has already started in \cite{barrow}
with quadratic gravity. The second one is the generalization to more
complicated metrics, which will allow us to consider different
physical framework. In the future, both of these thrusts will make
possible not only to develop experimental test for alternative
gravity but also to allow a better understanding of the reasons
underlying the success of General Relativity.

\end{document}